\documentclass[preprint,12pt]{elsarticle}
\usepackage{graphicx}
\usepackage{amssymb}
\biboptions{square,comma,sort&compress}
\usepackage{url}

\journal{Chemical Physics Letters}

\begin{document}

\begin{frontmatter}

\title{Optical spectrum of proflavine and its ions}

\author{A. Bonaca}
\ead{abonaca@gmail.com}

\author{G. Bilalbegovi\' c\corref{cor1}}
\ead{goranka.bilalbegovic@zg.t-com.hr}
\cortext[cor1]{Corresponding author}

\address{Department of Physics, Faculty of Science, University of Zagreb,\\
Bijeni{\v c}ka 32, 10000 Zagreb, Croatia}

\begin{abstract}
Motivated by possible astrophysical and biological applications
we calculate visible and near UV spectral lines of proflavine (C$_{13}$H$_{11}$N$_3$, 3,6-diaminoacridine) in vacuum, 
as well as its anion, cation, and dication.
The pseudopotential density functional and time-dependent density functional methods are used.
We find a good agreement in spectral line positions calculated by two real-time propagation methods and
the Lanczos chain method. Spectra of proflavine and its ions show characteristic UV lines which are good
candidates for a detection of these molecules in interstellar space and
various biological processes.
\end{abstract}

\begin{keyword}
Astrophysics \sep Astrochemistry \sep Biology \sep Molecular data \sep Density functional theory
\sep Time-dependent density functional theory

\end{keyword}

\end{frontmatter}

\section{Introduction}
\label{intro}
More than 150 molecules have already been detected in interstellar space \cite{Snow}.
The number of atoms in by now discovered interstellar molecules is in the range from two to more than ten.
Broad absorption spectral structures have been observed in the near-UV, visible, and near-IR spectra of some stars. They are called diffuse interstellar bands (DIBs) \cite{Snow,Herbig,Sarre,IAU}. Up to now, around 300 DIBs have been measured. Although the first such lines were detected almost ninety years ago, their carriers are still unknown. Many DIBs
carriers have been suggested, but two main groups are processes on the astrophysical dust grains and molecular transitions. The molecular origin of DIBs is more dominant hypothesis in the recent literature.  Various simple or complex carbon containing molecules have been suggested as possible DIBs carriers.
For example, small linear carbon chains, fullerenes, and especially
polycyclic aromatic hydrocarbons (PAHs)  have been investigated as candidates
for DIBs carriers \cite{IAU,Tielens,Malloci,Mulas,Weisman}.
Although various neutral and ionized PAHs, small and large ones,  have been extensively studied,
the analysis still did not unambiguously confirm agreement between any specific PAHs spectra and DIBs.
A few narrow DIBs have been observed in the Red Rectangle nebula \cite{Miles}.
Substructures in prominent bands of some narrow DIBs have been seen \cite{Kerr,Ehrenfreund}.
Features in these resolved profiles are consistent with rovibrational transitions in organic molecules.
It is also important to study various molecules  where atoms of other chemical elements substitute C and H atoms in PAHs. Nitrogen is a very important component of biological materials, and it is one of the most abundant chemical elements in the Universe.
Several molecules that contain nitrogen have already been detected in  meteorites, interstellar dust particles, and Space, or they have been analyzed by computational methods \cite{Clemett,Bauschlicher,Hudgins}.

Proflavine, C$_{13}$H$_{11}$N$_3$  (3,6-diaminoacridine), is a molecule
well known in molecular biology \cite{Lerman,Schulman,Jain}. This molecule interacts with nucleic acids, intercalates into DNA, and blocks activities of many  viruses. C$_{13}$H$_{11}$N$_3$  is a typical example of
amino-acridines which have been tested and used as anticancer drugs.
It has been suggested that proflavine is one of molecules which may act as ``midwifes'' \cite{Jain}.
It is possible that
this type of molecules increased a probability of DNA and RNA synthesis when biological matter was formed.
For example, proflavine increases thousand times a kinetics of the synthesis of two small DNA or RNA molecules into larger one. Therefore, proflavine is one of the candidates for molecules which induced the origin of life on Earth \cite{Thaddeus}. Optical spectra present an efficient way to detect proflavine, its ions, and derived nanostructures in biological materials and processes. 
In a comparison with PAHs (for example oligoacenes), available experimental and theoretical data for proflavine are scarce.
In ref. 3 it has been suggested that dyes are good candidates 
for studies of a DIBs problem and, as a support of this idea, it was stated that proflavine absorbs at 445 nm in water at pH 7. 
However, existing measurements of proflavine spectra are done in aqueous and other solutions \cite{Schulman,Mataga,Neumann,deSilvestri,Pileni}. 
It has been found that the optical spectrum of proflavine is very sensitive to the type of a solvent and pH factor \cite{Mataga,deSilvestri}.
There is a possibility that spectra of proflavine and its ions in vacuum and under astrophysical conditions are substantially different from ones in solutions.
In addition, it is known that many materials from  the group of organic dyes show
strong lines in a visible spectrum. For example, absorption lines of basic dyes such as methylene blue, thionine, neutral red, safranine, acridine orange, acridine yellow, and proflavine are all in an interval from 400 \AA\, to 800 \AA\, \cite{Neumann}. Although many organic dyes are good candidates for DIBs studies, because of a ``midwife'' idea \cite{Jain} and possible applications in astrobiology,  
we have selected proflavine for our study.

It is known that many molecules in interstellar space exist in its ionized state. 
It has been found that positive and negative ions of PAH molecules play an important role in the chemistry of diffuse interstellar clouds \cite{Lepp}.
Therefore,
it is also important to analyze ions of proflavine.  We use pseudopotential density functional (DFT) methods to compute the ground-state properties of proflavine and its ions, as well as the pseudopotential time-dependent density functional theory (TDDFT) to calculate their optical spectra. Pseudopotential methods are computationally less demanding than all-electrons methods and offer a possibility to study larger systems.
Methods based on the pseudopotential DFT are now substantially developed
to provide reliable results for structural, electronic and optical properties of various materials \cite{Martin}. TDDFT \cite{Runge} is recently used to describe optical properties of many materials \cite{Yabana,Malloci,Mulas,Weisman,Lopez,Varsano,Rocca}.
It is known that electronic spectra of PAHs are in the UV and visible part of the spectrum.
Optical spectra of PAHs and many other organic molecules in this region of the spectrum are well described by TDDFT \cite{Malloci,Mulas,Weisman,Lopez,Varsano,Rocca}.

\section{Computational methods}
\label{methods}

The ground-states of all proflavine related molecules are optimized independently
using the Quantum ESPRESSO program package
\cite{Giannozzi}.
For a majority of optical spectra
we use a real-time propagation method within TDDFT as implemented in the Octopus code \cite{Yabana,Octopus}.
The first step is to
calculate the ground state. Then, the system is perturbed by a small time-dependent potential
$\delta V_{ext}(\vec{r},t)=-\kappa z\delta t$.
This perturbation is turned off, and the time-evolution
of the system is followed.  If the dipole momentum of the system is
$\vec{d}(t)=-e\int \rho(\vec{r},t)\vec{r}d\vec{r}$, then its dynamical polarizability is obtained from
$\alpha(\omega)=\frac{1}{\kappa}\int e^{i\omega t}[\vec{d}(t)-\vec{d}(0)]dt$. Optical spectrum is calculated from
\begin{equation}
S(\omega)=\frac{2\omega}{\pi}Im \; \alpha(\omega).
\label{cross}
\end{equation}
The real space method and the spacing of 0.13 \AA{} are used. The simulation box is formed by adding spheres of the 4 \AA{} radii around each atom. The time-dependent equations are
integrated using the time step of $\delta t=0.0012$ $\hbar$/eV. The time evolution of
$\sim$ 7.9 fs is followed. The approximated enforced time-reversal symmetry algorithm
is applied \cite{Castro}.

We use the TDLDA approximation and the Perdew-Zunger exchange-correlation functional \cite{Perdew}
because of its good numerical stability in the pseudopotential TDDFT method.
The Troullier-Martins pseudopotentials are used \cite{tm}.
It is known that hybrid functionals in all-electron DFT methods produce results  which are  often in better agreement with experiments
than those obtained using LDA and GGA \cite{Koch}.
In contrast to all-electron codes, applications of hybrid functionals in the plane wave pseudopotential DFT methods are still in development.  This is even more pronounced for pseudopotential TDDFT methods.
We compare structural parameters of neutral proflavine  calculated in LDA with those obtained using the BLYP functional in the Quantum ESPRESSO \cite{Becke,Yang,Tozzini}.
We also calculate  optical properties of neutral proflavine using
the B3LYP functional in the latest version of the Octopus code, where hybrid functionals are still in development, and find an agreement between LDA and B3LYP optical spectra.
However, because of much better numerical stability, as well as availability and reliability of corresponding pseudopotentials, we present results for the TDLDA functional.
It has been shown that the TDLDA pseudopotential approximation in a time-dependent evolution  gives optical properties in agreement with experiments and other calculations, even for complex biological molecules
such as the green fluorescent protein  \cite{Lopez} and DNA bases \cite{Varsano}.

We also compare optical spectrum of proflavine calculated by the Octopus code with one obtained by the TDDFT module \cite{Walker,Rocca} of Quantum ESPRESSO \cite{Giannozzi}.
This module presents a new algorithm for the time dependent density functional perturbation theory. A susceptibility is determined as an off-diagonal matrix element of the resolvent of the Liouvillian superoperator. Computations are done using a Lanczos continued
fraction method. The optical spectrum of proflavine is calculated using the ground state computed in the Quantum ESPRESSO package. Periodic boundary conditions and the tetragonal simulation cell of sides length
30 a.u. and 16.7 a.u. are used. A local density  Perdew-Zunger  exchange-correlation  functional \cite{Perdew}, and ultrasoft pseudopotentials of  the RRKJ type \cite{Rappe} are applied. The kinetic energy cutoff of 25 Ry for wavefunctions, and 200 Ry for the charge density are used. The convergence properties in the Lanczos method are checked and the optical spectrum for 3500 recursion steps, and  with the broadening of 0.01 Ry is shown in this work.

\section{Results and Discussion}
\label{results}

\begin{figure}
\includegraphics[scale=0.4]{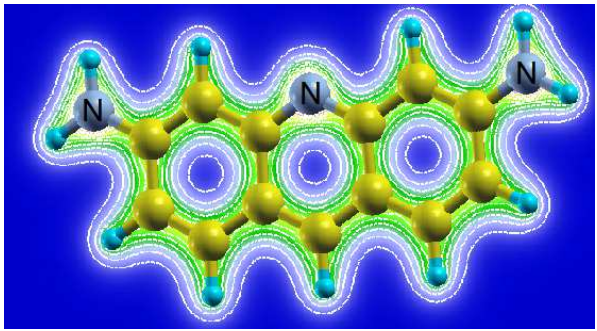}
\caption{
The structure and  electron density in the plane of a neutral proflavine molecule. Three nitrogen atoms are labeled by ``N''. Large balls
represent carbon atoms, whereas small ones are hydrogen atoms.
Isosurfaces are shown for
0.3 $e/a_0^3$ isovalues, where $a_0$  is the Bohr radius, and $e$ is the electron charge.
The figure is prepared using the XCrySDen package \cite{Kokalj}.}
\label{fig1}
\end{figure}

The structure and electron density of proflavine molecule is shown in Fig. \ref{fig1}.
It is found that different exchange functionals and corresponding pseudopotentials produce
results for geometry parameters which are sufficiently close to each other.
For example, in proflavine molecule average N-C distances are 1.365 \AA\, for LDA and norm-conserving pseudopotentials \cite{tm}, 1.349 \AA\, for LDA and ultra-soft pseudopotentials \cite{Rappe},
1.371 \AA\, for the BLYP functional \cite{Becke,Yang,Tozzini}.
Average C-C distances are: 1.395 \AA\, for  LDA and norm-conserving pseudopotentials,
1.396 \AA\, for  LDA and ultra-soft pseudopotentials, 1.413 \AA\, for the BLYP functional.
This should be compared with computationally generated data available in the NIST database \cite{NIST}:
1.365 \AA\, for average N-C, and 1.406 \AA\, for average C-C distances.

It is calculated that
the LDA Kohn-Sham HOMO-LUMO gap for neutral proflavine is 2.240 eV.
The quasiparticle corrected HOMO-LUMO gap is obtained using two formulae
\cite{Mulas,Jones,Godby}. From
$QP_{gap}^1 = E_{N+1} +E_{N-1} -2E_N$, where $E_N$ is the total energy of N-electron system,
the value of 4.509 eV is obtained.
The second formula for the quasiparticle corrected HOMO-LUMO gap
$QP_{gap}^2 = \epsilon _{N+1}^{N+1} - \epsilon _{N}^{N} $, where $\epsilon _{i}^{j}$ is the
i-th eigenvalue of the j-electron system, produces the value of 4.511 eV.

\begin{figure}
\includegraphics[scale=0.65]{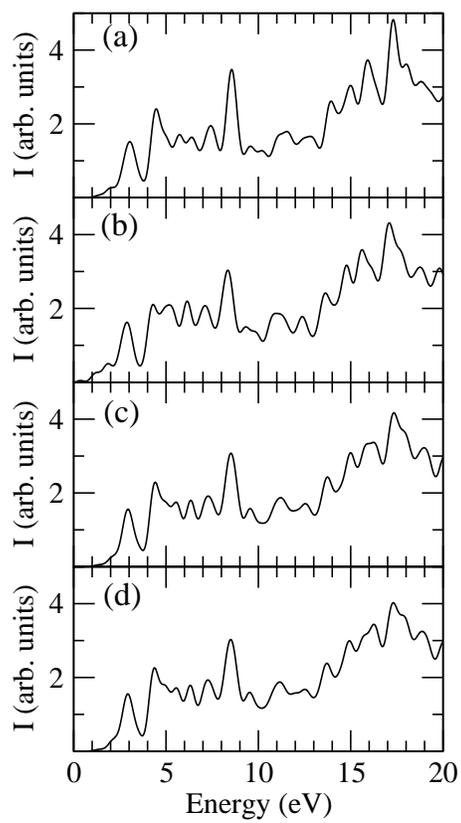}
\caption{Optical absorption spectra: (a) C$_{13}$H$_{11}$N$_3$, (b)  C$_{13}$H$_{11}$N$_3$$^{-1}$,
(c) C$_{13}$H$_{11}$N$_3$$^{+1}$, (d) C$_{13}$H$_{11}$N$_3$$^{+2}$.}
\label{fig2}
\end{figure}

\begin{figure}
\includegraphics[scale=0.55]{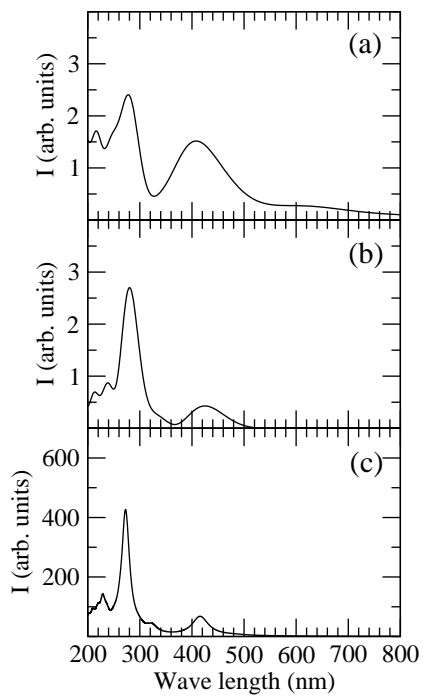}
\caption{Spectral lines of neutral proflavine in the near
UV ($\lambda _1$) and visible ($\lambda _2$) region calculated by three pseudopotential TDDFT methods:
(a) the TDLDA real-time propagation method  \cite{Octopus}, $\lambda _1=277.37$ nm, $\lambda _2=407.84$ nm,
(b) the TDB3LYP real-time propagation method   \cite{Octopus}, $\lambda _1= 280.54$ nm,
$\lambda _2=424.66$ nm,
(c) the Lanczos chains method \cite{Walker,Rocca,Giannozzi}, $\lambda _1= 271.76$ nm, $\lambda _2= 412.76$ nm.}
\label{fig3}
\end{figure}

Optical spectra of proflavine and its ions are shown in Fig. \ref{fig2}.
It is found that the overall spectra are rather similar to each other.
However, differences in the position and shape of lines exist. Optical transitions in ions occur
at lower energies than in a neutral proflavine molecule. The same relation between spectra has been found for  neutral and charged PAH molecules \cite{Hirata}.
We find that the first pronounced absorption peak in spectra of proflavine and its ions
appears at $\sim 3$ eV. The first absorption peak
in PAH anthracene (with the same number of cycles) is at 5 eV \cite{Mulas}.
Figure \ref{fig3} compares spectra of neutral proflavine calculated by two exchange-correlation functionals in the Octopus code \cite{Octopus} with the same result obtained by the TDDFT module \cite{Walker,Rocca} of Quantum Espresso \cite{Giannozzi}.
Although numerical algorithms and pseudopotentials used in these methods are different,
the agreement between shapes of the spectra and positions of lines is good.
It has been measured that a position of a visible absorption line of  proflavine in water at room temperature
changes from 444 nm to 394 nm when pH changes between 7.0 and 14.0 \cite{deSilvestri}.
These authors concluded that the value of 394 nm corresponds to the neutral form of proflavine in water.
The same line is at 397 nm for proflavine in purified toluene \cite{deSilvestri}.
The value for the maximum of the absorption spectrum of proflavine in water reported in ref. \cite{Pileni} is 400 nm.

The comparison of calculated lines and corresponding DIB lines \cite{Jenniskens,York,Tuairisg} is presented in Table \ref{pfDIB}. Data for a visible spectrum show that only line of a neutral molecule is within FWHM of known DIB lines.
Positions of the UV line nearest to the visible spectrum, as well as of a characteristic strong line which appears in all spectra between 8.35 eV and 8.55 eV, are also shown in Table \ref{pfDIB}.
Many organic molecules exhibit absorption lines in the UV  spectrum.  Therefore, the measurements and identification of UV spectral features are important for organic materials in Space.
For example,
the interval between 1320 \AA\, and 1440 \AA\, in the UV spectrum measured by the Space Telescope Imaging Spectrograph (STIS) has been analyzed \cite{Destree}.
The Cosmic Origins Spectrograph (COS), recently installed on the Hubble Space Telescope, will provide new data for spectra of organic molecules in UV \cite{IAU}.
Calculations based on a model of the state of hydrogenation and the degree of ionization in the diffuse interstellar medium 
have shown that small PAHs with 15-20 carbon atoms are destroyed in the UV field \cite{LePage}.
Although proflavine and its ions are rather small, there is a possibility that they exist in Space adsorbed on suitable grains, or as intercalators in  some bigger molecules. That could even change their optical spectra and give line positions in better agreement with DIBs.

\begin{table}
\centering{
\caption{Comparison of calculated optical spectral lines of proflavine and its ions with the closest known DIB lines \cite{Jenniskens,York,Tuairisg}. FWHM of measured lines is also shown, as well as positions of two UV lines to facilitate a comparison with future  measurements of the ultraviolet range in spectra
of interstellar organics.}
\begin{tabular}{l c c c}
\hline
Molecule & Lines (\AA{}) & DIB (\AA{}) & FWHM (\AA{}) \\
\hline
C$_{13}$H$_{11}$N$_3$ & 4078.42 & 4066.0 \cite{Jenniskens}& 15.0 \\
& 2773.69 & &\\
& 1450.11 & &\\
C$_{13}$H$_{11}$N$_3$$^{-1}$ & 4305.00 & 4363.86 \cite{York}& 0.46 \\
	& 2883.35&  &\\
	& 1484.84 &  & \\
C$_{13}$H$_{11}$N$_3$$^{+1}$ & 4217.15 & 4175.46 \cite{Tuairisg} & 17.2 \\
	& 2805.07 &  &\\
	&1455.22 & & \\
C$_{13}$H$_{11}$N$_3$$^{+2}$ & 4232.08 & 4259.01 \cite{York} & 1.05\\
	& 2831.05 &  &\\
	& 1458.64 & & \\	
\hline	
\end{tabular}
\label{pfDIB}
}
\end{table}

\section{Conclusions}
\label{concl}

We calculated optical spectra of proflavine, its
anion, cation, and dication using pseudopotential density functional theory methods.
A comparison of three TDDFT pseudopotential methods shows that they produce spectra of proflavine which  qualitatively agree well, and line positions differ 0.14 eV, at the most.
The difference of an average result (calculated by three TDDFT methods) for a visible line position and a corresponding average experimental
value for a neutral proflavine molecule in water \cite{deSilvestri,Pileni} is also 0.14 eV.
The shapes of spectra for proflavine and its ions are rather similar to each other, but the change in the
line positions induced by a different charge is apparent.
UV lines are characteristic for spectra of proflavine and its ions and give an efficient
way for their detection, either in Space, meteorites, astrophysical
dust particles, or in biological processes. However, the UV spectral region is much less studied than visible one.
It is important to analyze available data and to perform new studies of substituted polycyclic aromatic hydrocarbons.
We hope that our work will generate an interest for measurements of  optical spectra of proflavine and other organic dyes in the gaseous phase and under various astrophysical conditions.

\section*{Acknowledgement}
We are grateful to the HR-MZOS project ``Electronic Properties of Surfaces and Nanostructures'' for support of this research and the University of Zagreb Computing Center SRCE for providing computer resources.
We would like to thank  Stefano Baroni, Alberto Castro, Osman Baris Malcioglu and Layla Martin-Samos for
discussions.

\end{document}